\documentclass[letterpaper,11pt]{article}
\pdfoutput=1
\usepackage{jheppub} 
\usepackage{slashed}
\usepackage{graphicx}
\usepackage{subfigure}
\usepackage{soul}
\usepackage{amsmath}
\usepackage{multirow}
\usepackage{tablefootnote}
\usepackage{comment}
\usepackage{hyperref} 
\usepackage{epstopdf} 
\usepackage{color}
\usepackage{physics}

\newcommand{\beq}{\begin{equation}}
\newcommand{\eeq}{\end{equation}}
\newcommand{\bea}{\begin{eqnarray}}
\newcommand{\eea}{\end{eqnarray}}

\def\m1{M_1}
\def\m2{M_2}
\def\m3{M_3}

\def\ch10{\tilde \chi^0_1}

\def\gev{\,{\rm GeV}}

\def\to{\rightarrow}

\newcommand{\lsim}{\mathrel{\mathop{\kern 0pt \rlap
  {\raise.2ex\hbox{$<$}}}
  \lower.9ex\hbox{\kern-.190em $\sim$}}}
\newcommand{\gsim}{\mathrel{\mathop{\kern 0pt \rlap
  {\raise.2ex\hbox{$>$}}}
  \lower.9ex\hbox{\kern-.190em $\sim$}}}

\definecolor{pink}{RGB}{255,105,180}

\title{Would a Deeply Bound $b\bar b b\bar b$ Tetraquark Meson be Observed at the LHC?}

\author{Estia Eichten, }
\author{Zhen Liu}

\affiliation{Theoretical Physics Department, Fermi National Accelerator Laboratory, Batavia, IL, 60510}

\emailAdd{eichten@fnal.gov}
\emailAdd{zliu2@fnal.gov}

\abstract{

There has been much theoretical speculation about the existence of a deeply bounded tetra-bottom state.  Such a
state would not be expected to be more than a GeV below $\Upsilon\Upsilon$ threshold.  If such a state exists below 
the $\eta_b\eta_b$ threshold it would be narrow,  as Zweig allowed strong decays are kinematically forbidden.  
Given the observation  of $\Upsilon$ pair production at CMS,  such a state with a large branching fraction into $\Upsilon \Upsilon^*$ is likely 
discoverable at the LHC.  The discovery mode is similar to the SM Higgs decaying into four leptons through the $Z Z^*$ channel.  
The testable features of both production and the four lepton decays of such a tetra-bottom ground state are presented.
The assumptions required for each feature are identified,  allowing the application of our results more generally to a resonance decaying 
into four charged leptons (through the $\Upsilon\Upsilon^*$ channel)  in the same mass region.}


\preprint{
\begin{flushright}
FERMILAB-PUB-17-395-T
\end{flushright}
}

\begin{document}
\maketitle
\flushbottom

\section{Introduction}

Since the discovery of the X(3872) \cite{Choi:2003ue} the possibility of meson states with four valence quarks has received considerable attention.   Many other quarkonium-like states, the so-called XYZ states \cite{Agashe:2014kda} have since been observed.  Theoretical models have been proposed to explain these states involving systems involving a heavy quark-antiquark pair  (c or b) and a light quark-antiquark pair (u, d, s) \cite{Brambilla:2010cs, Bodwin:2013nua} .   In particular, the discovery of isospin one resonances with hidden heavy flavor quarks,  the $Z_c^{\pm}(3900), Z_c^{\pm}(4020)$ 
and $Z_b^{\pm}(10610), Z_b^{\pm}(10650)$  states \cite{Patrignani:2016xqp} makes the interpretation of all these states without additional light valence quarks impossible.  

In all the presently observed XYZ states, the tetraquark state is very near or above the threshold for strong decays (Zweig allowed) 
into a pair of heavy-light mesons.  
It is therefore natural to ask what happens as the mass of the lighter quarks is raised so that all four quarks become heavy.  
Could the binding become stronger as the mass increases as is observed for heavy quark-antiquark (quarkonium) systems?  
This could lead to narrow deeply bound tetraquark systems (without Zweig allowed strong decays to quarkonium states).  There is some theoretical reasons to suggest this maybe the case.  In the QED analog, the lowest state of two positronium 
atoms $P_{s2}$ (a positronium molecule) is bound~\cite{Wheeler:1946, Hyllerass:1947, Varga:1998} and has been unambiguously observed  in 2007~\cite{Cassidy:2007}.  Similarly in the perturbative NRQCD limit  of four heavy quarks, the Van der Waals force between two color singlet mesons separated by large 
distance is attractive~\cite{Brambilla:2017ffe}.  

The heaviest tetraquark system involve four bottom flavored quarks.   If the mass of the lowest such tetraquark state were below 
$\eta_b\eta_b$ threshold,  the decays would occur only by the annihilation of one quark-antiquark pair and the state will be very narrow.  
In the following sections this possibility and its consequences for observation at the LHC are explored in detail.

\section{States with four heavy quarks}

If all four quarks are  heavy, we may use NRQCD to study these systems.  There are three approaches to study these systems: (1) Direct measurement of the spectrum using Lattice QCD, (2) QCD sum rule approach, and (3) Non relativistic potential models motivated by QCD expectations.  
Direct lattice calculations should be very informative but have not yet been done.   If the ground state of the four quark system would be significantly below the threshold of pair production of two quarkonium states in the same $J^{PC}$ channel,  the required lattice calculations is greatly simplified.   
Some calculations using the QCD sum rule approach have been presented recently~\cite{Chen:2016jxd,Wang:2017jtz}. 
They conclude that for the 4 b quark system the $J^{PC}$ ground states are below the strong decay threshold but for the 4 c quark systems all the states are above thresholds for strong decay.  The third approach of using QCD inspired potential models will be discussed below.

The Hamiltonian, H, for four heavy quarks is the sum of the non relativistic kinetic energy, T, and a potential energy, V, which expresses the interactions between the heavy quarks,
\begin{equation}
H =  T_{kin} + V.
\end{equation}
Consider two quarks $Q_1, Q_3$  at positions $\vec r_1, \vec r_3$  and antiquarks $\bar Q_2, \bar Q_4$  at positions $\vec r_2, \vec r_4$ respectively.  In the non relativistic limit the quark spin can be treated as a relativistic correction.  The overall position, $\vec R = \sum_i \vec r_i$, and angular momentum, L,  separate as with the usual two-body Schr\"{o}dinger equation.  However, we are left with six variables and a very complicated Hamiltonian to solve for the energies and wavefuctions of the various states.  This Hamiltonian can only be solved numerically,  so we are limited here to present some general remarks about the form.  

\begin{figure}[htbp]
   \centering
   \includegraphics[scale=0.5,clip]{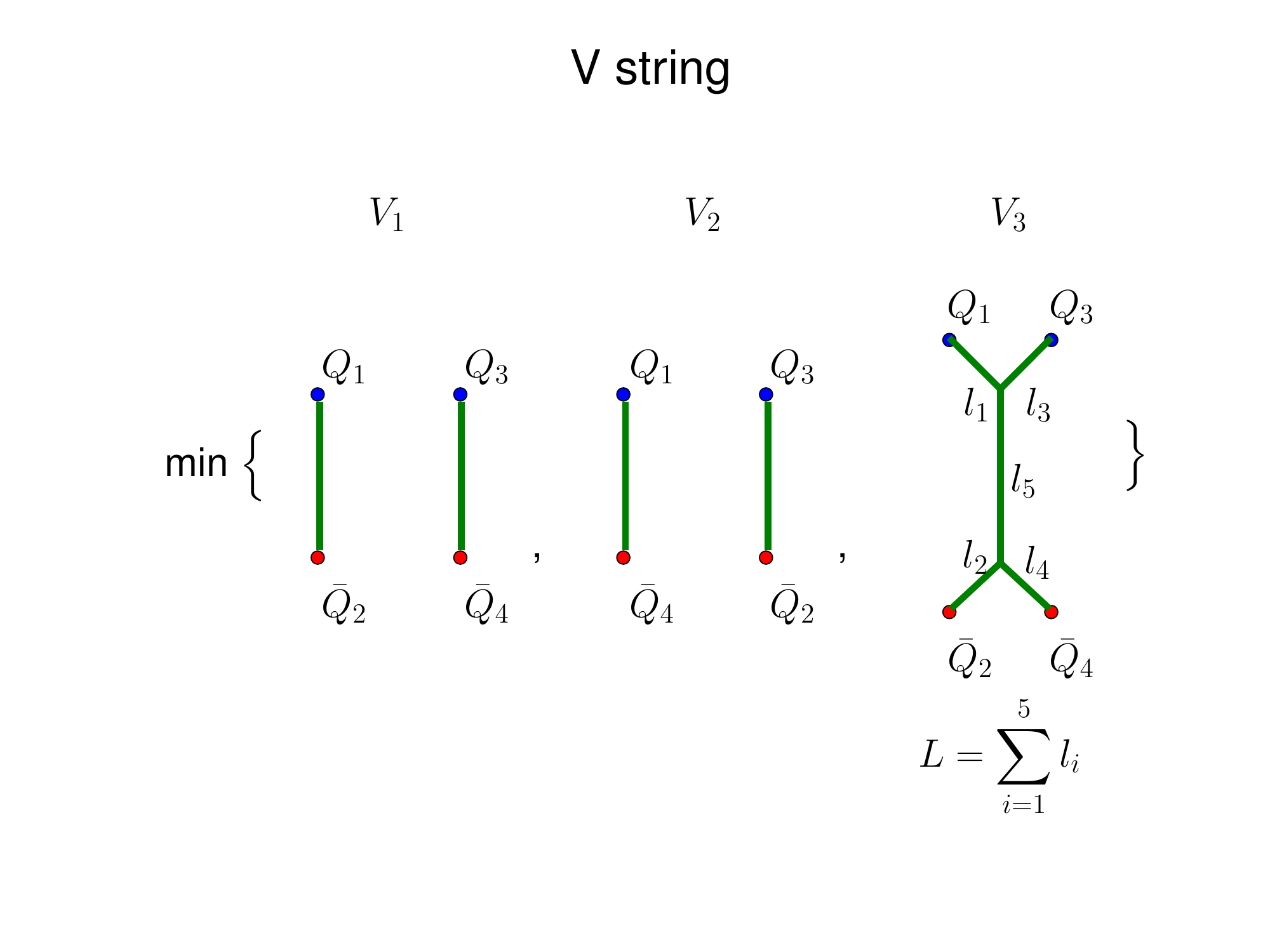}
   \caption{The long range tetraquark potential.}
   \label{fig:pot}
\end{figure}

Denote the relative distances between the four quarks by the six values $r_{ij} = |\vec r_i - \vec r_j|$ for $i<j$ the short distance behaviour of the potential is given by perturbative QCD.
In lowest order 
\begin{equation}
V = V_{pQCD} + V_{string}.
\end{equation}
$V_{pQCD}$ is the perturbative one gluon exchange terms of the form
\begin{equation}
V_{pQCD}  =   \sum_{i,j ~for~ i<j} c_{ij} \frac{\alpha_s(r_{ij}) }{r_{ij}}
\end{equation}
where $c_{ij}$ are the SU(3) Clebsch-Gordon coefficients for single gluon exchange.  In the limit that all quark masses are extremely large the ground 
state is determined by perturbative QCD alone.  However,  not even the b quark is sufficiently heavy to ignore the non perturbative QCD interactions.
The long distance part  $V_{string}$ is modeled by the string terms as shown in Figure \ref{fig:pot}.
It is determined by the shortest path that creates a local color singlet state.  $V_1 = \sigma  (r_{12}+r_{34})$,  $V_2= \sigma (r_{14} + r_{32})$ and $V_3 = \sum_i (\epsilon (1,2,i)\epsilon(i, 4, 3) ) \sigma L$ with L being the length of the shortest path that couples all the quarks (see Fig \ref{fig:pot}).  The string tension is denoted $\sigma$. Finally $V_{string} = min (V_1, V_2, V_3)$.  This form has the interesting behaviour of flipping from one form to another as relative distances change.  The form of this potential is consistent with recent lattice studies of the tetraquark static potential~\cite{Bicudo:2017usw}.

Unlike the usual mesons and baryons, the  tetraquark system has two separate color singlet combinations: 
$(3\times \bar 3\times \bar 3 \times 3)$ can be decomposed as $\rightarrow  (1 \times 1) + (8 \times 8)$ [or an alternately basis $ (\bar 3 \times 3) +(6 \times \bar 6) $],  
i.e. two unique ways to get a singlet.  Thus the wavefunction of the tetraquark states have two components in color space. 
If quarks (1,2,3,4) have colors indices (i,j,k,l) we can chose the basis  ($\delta^i_j \delta^k_l$)  for $\psi_I$ and  ($\delta^i_l \delta^k_j$)  for $\psi_{II}$
to represent these two components.  So properly the potential is a $2\times2$ matrix in color space.  Thus $V_{pQCD}$ is given by
\begin{equation}
 \left \{ \begin{array}{cc}   - \frac{4}{3} (v(r_{13}) + v(r_{24}))  & \frac{4}{9} (v(r_{12})+ v(r_{34}))  \\ 
\frac{4}{9}(v(r_{12})+ v(r_{34}))  & - \frac{4}{3} (v(r_{14}) + v(r_{23})) 
\end{array} \right \}
\label{eq:Vp}
\end{equation}
with $v(r_{ij}) \equiv \alpha(r_{ij})/r_{ij}$.

In a similar way $V_{string}$ can be written in the form:
\begin{equation}
 V_{string}  \left [ \begin{array}{c}\psi_I \\ \psi_{II} \end{array}\right ] = \left \{ \begin{array}{cc}   V_1+  \beta_{11} V_3 & - \beta_{12}V_3 \\ 
 -\beta_{21}V_3 &  V_2 + \beta_{22} V_3 \end{array} \right \}
  \left [ \begin{array}{c}\psi_I \\ \psi_{II} \end{array}\right ] 
\label{eq:Vs}
\end{equation}
where $\beta_{ij}$ is the matrix projection of the $V_3$ potential on the two color states. 

In the various limits the expected behaviour is recovered. For $r_{13}$ and $r_{24}$ fixed and all the other distances becoming large, the solutions decompose into the two 
mesons $A={Q_1\bar Q_3}$ and  $B={Q_2\bar Q_4}$ and  $\psi_I  \approx  \psi_A(r_{13}) \psi_B(r_{34}) $.  Similarly with $r_{14}$ and $r_{23}$ fixed and other distances large
the the solutions decompose  into two mesons $A={Q_1\bar Q_4}$ and  $B={Q_2\bar Q_3}$.   Notice that in both of these cases the resulting Hamiltonian is just the usual potential for quarkonium states A and B.  In either case it is useful to decompose the kinetic energy of the reduced system 
\begin{equation}
T_{kin} = - (\frac{1}{2\mu_A} \vec \nabla_{r_A}^2  + \frac{1}{2\mu_B} \vec \nabla_{r_B}^2 + \frac{1}{ 2\mu_{AB}} \vec \nabla_{r_{AB}}^2)  \left \{ \begin{array} {cc} 1 & 0 \\ 0 & 1\end{array} \right \} 
\end{equation}
where $\vec r_A$ and $\vec r_B$ are the relative position of the quark and antiquark in meson A and B respectively; $\vec r_{AB}$ is the relative position of the center of masses of
meson $A$ and $B$; and $\mu_A, \mu_B$ and $\mu_{AB}$ are the associated reduced masses of the subsystems.~\footnote{For example, for  $A={Q_1\bar Q_3}$ and  $B={Q_2\bar Q_4}$,
$\mu_A = \frac{m_1 m_3}{m_1+m_3}$,  $\mu_B =\frac{m_2 m_4}{m_2+m_4}$, $\mu_{AB} = \frac{(m_1+m_3)(m_2+m_4)}{m_1+m_2+m_3+m_4}$, 
$\vec r_A =  \mu_A (\frac{\vec r_1}{m_3} -\frac{\vec r_3}{m_1})$, $\vec r_B=  \mu_B (\frac{\vec r_2}{m_4} -\frac{\vec r_4}{m_2})$, and $r_{AB} =\mu_{AB}(\frac{\vec r_1}{m_3} -\frac{\vec r_3}{m_1})$.}  

In the limit $r_{12}$ and $r_{34}$ fixed and the other distances becoming large, the diquark-antidiquark system is approximated. Here the dynamics is separated into 
the binding of the diquark $A={Q_1 Q_2}$ and antidiquark $B={\bar Q_3 \bar Q_4}$ systems into $\bar 3$ and $3$ systems.  These systems then are bound in the overall singlet state just like a quarkonium system.  The wavefunction for the diquark-antidiquark state ($\bar 3 \times 3$) is simply  $\sqrt{\frac{1}{2}} (\psi_I - \psi_{II})$.
Note that the  $6$ and $\bar 6$ systems will not be relevant for low-lying states because the short range piece of $V_{pQCD}$ is repulsive and the lowest order long range string
potential $V_3$ requires the diquark and antidiquark to be $\bar 3$ and $3$ respectively.  
In general the full spectrum of systems with four heavy quarks has not yet been calculated.  Even the dominate spatial contributions to the wavefunction of the ground state system  remains unresolved.

Under various assumptions tetraquark systems have been studied.  Detailed studies of within the Bethe-Salpeter approach has been presented by 
Heupel, Eichmann and Fischer~\cite{Heupel:2012ua, Eichmann:2015nra, Eichmann:2015zwa} for tetraquark systems with lighter quark masses (up to the charm quark mass).  
However, only the lowest order QCD one gluon exchanges are included in the kernel at present.  In the limit of sufficiently heavy 
quarks the inclusion of only the lowest order gluon exchanges would be would be rigorous.  In this heavy quark limit, ground state masses has been investigated using a variational technique by Czarnecki,  Leng and Voloshin\cite{Czarnecki:2017vco}. They conclude that such tetraquark systems with all equal masses are not bound.  
More phenomenological approaches in which it is assumed that dynamics of the tetraquark system is approximated by a diquark-antidiquark $(\bar 3\times 3)$
system have also been studied~\cite{Berezhnoy:2011xn}.  Here  narrow tetraquark states below threshold for strong decays are found for both $(c\bar c c\bar c)$ and $(b\bar b b\bar b)$ systems. Using the two-body subsystems for 4 heavy quarks the tetraquark spectrum has been studied~\cite{Popovici:2014usa}.
Bai, Lu and Osborne have studied the ground state of the $b\bar b b\bar b$ system including the non-perturbative string potential (Fig.\ref{fig:pot}) using 
the Diffusion Monte Carlo method \cite{Bai:2016int}.  They find the ground state $0^{++}$ tetra-b quark state is bound, while Richard, Valcarce and Vijande  argue that such states will not be bound~\cite{Richard:2017vry} and a phenomenological analysis of Karliner, Nussinov and Rosner puts this state just below di-$\Upsilon$ threshold \cite{Karliner:2016zzc}.
One can only conclude at present that the issue of binding awaits a definitive Lattice QCD calculation.   
We will discuss what can be said reliably in the next section.

\section{Phenomenology of the Low-lying $b\bar b b\bar b$ tetraquark states}

At leading order the spin-splittings can be ignored and states are described by its  radial quantum numbers and its orbital angular momentum $L = l_A + l_B + l_{AB}$.
The ground state  would be expected to be fully symmetric, so $L=0$ with the subsystem angular momenta  $l_A, l_B, l_{AB}$ all zero as well.  
After adding spin there are in general six degenerate states: $J^{PC} = 0^{++}, 0^{++'}, 1^{+-},1^{+-'},1^{++}, 2^{++}$.  
If the two quarks are identical as in the $(b\bar bb\bar b)$ system then ${Q_1 Q_2}$ will be antisymmetric in color,  
then for the ground state the total spin of the two quarks (two antiquarks) must also be symmetric $S=1$ state,  
hence the  tetraquark system can have $J^{PC} = 0^{++}, 1^{+-}, 2^{++}$.  In terms of the diquark-antidiquark basis
the states are shown in Table \ref{tab:states}.
\begin{table}[htbp]
   \centering
    \begin{tabular}{@{} c c c c  @{}} 
    \hline
      $J^{PC}$     & color structure & $ < J; s \bar s> $ & $<J; M_1 M_2> $\\
      \multicolumn{4}{c}{equal mass quarks} \\
      \hline
       $0^{++}$   & $(\bar 3 \times 3)$  & $|0;1,1>$   &  $ - \frac{1}{2} |0; V, V> + \frac{\sqrt{3}}{2} |0;P,P>$ \\
       $1^{+-}$    & $(\bar 3 \times 3)$  & $|1;1,1>$   & $ \frac{1}{\sqrt{2}} (|1;V,P> + |1;P,V>)$ \\
       $2{++}$     & $(\bar 3 \times 3)$  & $|2;1,1>$   &  $|2; V,V>$ \\ 
       $0^{''++}$   & $(6 \times \bar 6)$  & $|0;0,0>$  &  $\frac{\sqrt{3}}{2} |0; V, V> + \frac{1}{2} |0:P,P> $\\
       \hline
        \multicolumn{4}{c}{additional states for unequal mass quarks} \\
       $0^{'++}$   & $(\bar 3 \times 3)$  & $|0;0,0>$  & $\frac{\sqrt{3}}{2} |0; V, V> + \frac{1}{2} |0:P,P>$ \\
        $1^{++}$    & $(\bar 3 \times 3)$  & $\frac{1}{\sqrt{2}}(|1;1,0>+|1;0,1>)$ & $0$ \\ 
       $1^{'+-}$    & $(\bar 3 \times 3)$  & $\frac{1}{\sqrt{2}}(|1;1,0>-|1;0,1>)$ & $\frac{1}{\sqrt{2}} (|1; V,P> - |1;P,V>)$ \\ 
       $0^{'''++}$   & $(6 \times \bar 6)$  & $|0;1,1>$ &$ - \frac{1}{2} |0; V, V> + \frac{\sqrt{3}}{2} |0;P,P>$ \\
       $1^{''++}$   & $(6 \times \bar 6)$  & $|1;1,1>$  & $ \frac{1}{\sqrt{2}} (|1;V,P> + |1;P,V>)$\\
       $2^{'++}$   & $(6 \times \bar 6)$  & $|2;1,1>$  & $|2; V,V>$ \\
      \hline
   \end{tabular}
   \caption{Coupling coefficients of tetraquark $J^{PC}$ ground states (fully symmetric in space)  to quarkonium pseudoscalar (P) and vector (V) ground states.  Both color diquark combinations are shown.  For the $bb\bar b\bar b$ and $
 cc\bar c\bar c$ systems both color states [with anti-triplet diquark component with spin ($s = 1$) and  sextet component with spin ($s=0$)],  it is expected that the attractive $\bar 3\times 3$ channels are all lower states and the $6\times \bar 6$ channels are   excited states.  Since the tetraquark state is a color singlet,  the antidiquark component must have spin ($\bar s = s$).  For the unequal mass system,  $bc\bar b\bar c$, there are additional states as shown.}
   \label{tab:states}
\end{table}

\subsection{Decay properties}

\begin{table}[htbp]
   \centering
    \begin{tabular}{@{} c c c c c c c  @{}} 
    \hline
      $J^{PC}$     &  $s$ & $\bar s$ & P P & PV& VP & VV \\
      \multicolumn{7}{c}{equal mass quarks} \\
      \hline
       $0^{++}$   & $1$  & $1$ & $\sqrt{3}/2$ & $0$ & $0$ & $-1/2$  \\
       $1^{+-}$  &  $1$  & $1$ & $0$ &  $1/\sqrt{2}$ & $1/\sqrt{2}$ & $0$ \\ 
       $2{++}$   & $1$  & $1$ & $0$ & $0$ & $0$ & $1$  \\
       \hline
        \multicolumn{7}{c}{additional states for unequal mass quarks} \\
       $0^{++}$  &  $0$  & $0$ & $1/2$ &  $0$ & $0$ & $\sqrt{3}/2$ \\ 
       $1^{+-}$  &  $1$  & $0$ & $0$ &  $-1/2$ & $1/2$ & $1/\sqrt{2}$ \\ 
       $1^{++}$  &  $0$  & $1$ & $0$ &  $1/2$ &  $-1/2$ &$1/\sqrt{2}$ \\ 

      \hline
   \end{tabular}
   \caption{Coupling coefficients of tetraquark $J^{PC}$ ground states (fully symmetric in space)  to quarkonium pseudoscalar (P) and vector (V) ground states.  Both color diquark combinations are shown.  For the $bb\bar b\bar b$ and $
 cc\bar c\bar c$ systems both color states [with anti-triplet diquark component with spin ($s = 1$) and  sextet component with spin ($s=0$)] .  It is expected that the attractive $\bar 3\times 3$ channels are all lower and the $6\times \bar 6$ channels are excited states.  Since the tetraquark state is a color singlet,  the antidiquark component must have spin ($\bar s = s$). 
For the unequal mass system $bc\bar b\bar c$ additional states are allowed as shown.}
   \label{tab:couplings}
\end{table}

The decay property of this tetraquark state could be fully determined by the effective Lagrangian,
\beq
\Delta\mathcal{L} = \frac 1 2 (\partial_\mu \phi)^2 - \frac 1 2 \Lambda \phi \Upsilon^\mu \Upsilon_\mu + ...,
\label{eq:EFTs}
\eeq 
where the dimensionful quantity $\Lambda$ characterizes the interaction between the ground state $\phi$ and $\Upsilon\Upsilon$ states. 
In principle, $\Upsilon$ represents $\Upsilon$ 1S, 2S, 3S, etc., states and the coefficients $\Lambda$ could differ for different $\Upsilon$ state combinations.  We omit possible higher dimensional interaction terms (or a more general form factor) that feature different Lorentz structure as we anticipate them to generate  sub-leading contributions to the production and decays of the tetraquark state. Assuming the state $\phi$ is a deeply bounded state (below $\eta_b\eta_b$ threshold), it is not unreasonable to assume that the ground state dominates. 
Throughout this paper, unless otherwise noted, we only consider the state $\phi$ overlapping with the $\Upsilon(1S)$ through this basic interaction term.
There is no {\it priori} knowledge about the size of this parameter $\Lambda$.\footnote{The amplitude for the  strong decay of a tetraquark $(b\bar b b\bar b$) state into two $(b\bar b$) quarkonium
states would expected to be a typical QCD scale (i.e. $\Lambda \approx \Lambda_{\rm QCD} \approx$ 200 MeV).}

Similar to the $0^{++}$ case explained above, if the underlying state is a pseudoscalar, vector or a tensor, different forms of allowed Effective Field Theory (EFT) with di-Upsilon system are allowed. If such a state exists, the differential observables would help determine the structure of the  EFT and thus the associated $J^{PC}$ of the tetraquark state. We tabulate the possible states and operators in Table.~\ref{tab:EFTs}, keeping only the lowest dimensional operators. We anticipate leading observable state would be the tetra quark ground state $0^{++}$ and thus list a few other possibilities to contrast and check, we omit the possibility spin-1 or CP-odd spin-2 state for simplicity.

\begin{table}
  \centering
  \begin{tabular}{@{} c c c c c   @{}}
  \hline
  State & $J^{PC}$ & Leading Operator \\ \hline
   $\phi$ & $0^{++}$ & $\phi \Upsilon^\mu \Upsilon_\mu$ \\
   $\tilde \phi$ & $0^{-+}$ & $\tilde \phi \Upsilon^{\mu\nu} \tilde \Upsilon_{\mu\nu}$\\ 
   $T_{\mu\nu}$ & $2^{++}$ & $T_{\mu\nu}\Upsilon^{\mu} \Upsilon^{\nu}$\\ 
\hline
\end{tabular}
\caption{\label{tab:EFTs} Benchmark tetraquark states that could couple to di-Upsilon states along with the associated form of the leading interaction term.}
\end{table}

\begin{figure}[htbp]
  \begin{center}
  \includegraphics[scale=1.0,clip]{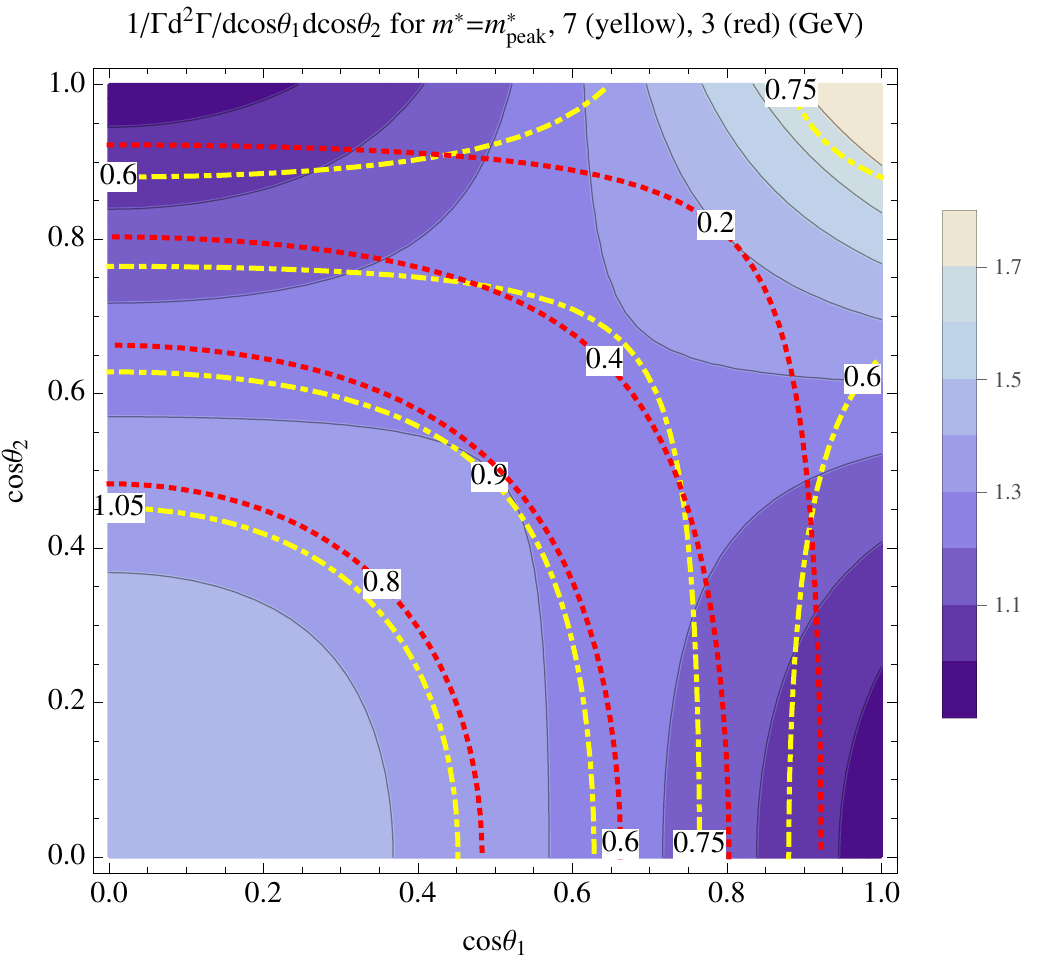}
  \includegraphics[scale=0.7,clip]{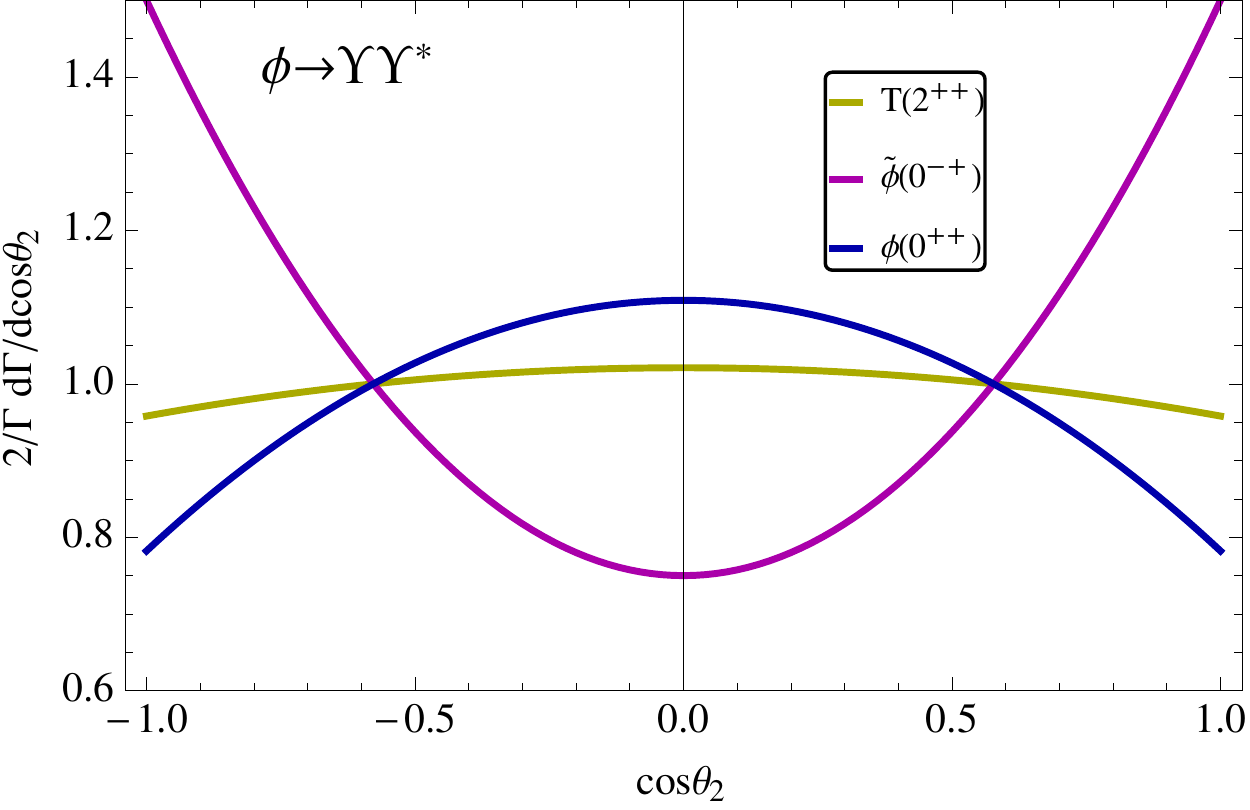}
  \caption{
Upper panel: double differential angular distribution of the tetraquark state $\phi(0^{++})$ for different values of the off-shell dilepton invariant masses. The distribution is symmetric for negative values of $\cos\theta_1$ and $\cos\theta_2$, the polar angles of the leptons in their dilepton rest frames, assuming $\phi$ being a $0^{++}$ state. 
Lower panel: the angular distributions (averaged over all invariant mass bins) in the rest frame of the dilepton pairs that from the same $\Upsilon^{(*)}$ 
for different assumptions of the spin and parities of the resonant particle.
  }
  \label{fig:decayangular}
  \end{center}
\end{figure}

Considering the observability at the hadron collider environment, we focus on the purely leptonic decays of the tetraquark state. These purely leptonic decays can be understood as mediated by the intermediate (on-shell or off-shell) vector meson states $\Upsilon\Upsilon^*$.
Since the $\Upsilon$ pair production has been measured directly at both experiment with low background, such a resonant four lepton final state will be also observable if produced with sufficient rate. The production properties for this possible deeply bounded tetraquark state will be discussed in the next section. 

\begin{figure}[htbp]
  \begin{center}
  \includegraphics[scale=0.7,clip]{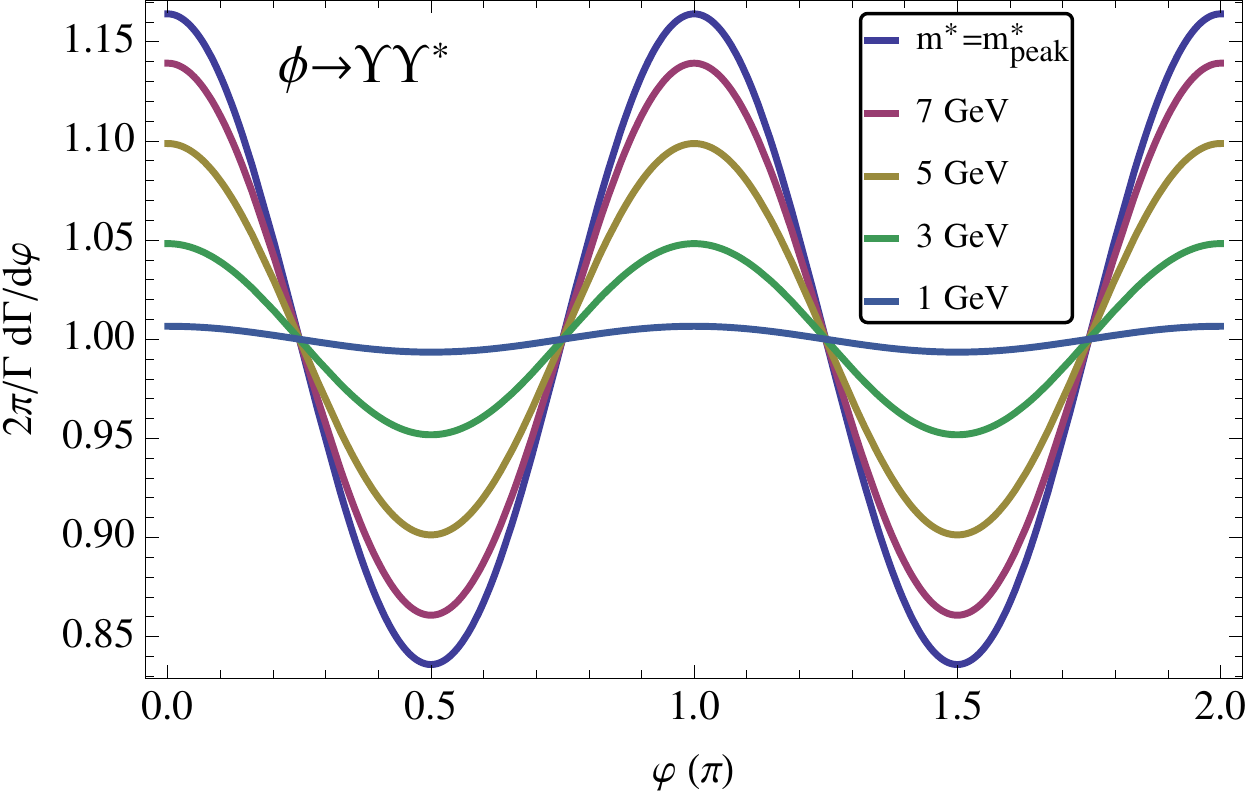}
  \includegraphics[scale=0.7,clip]{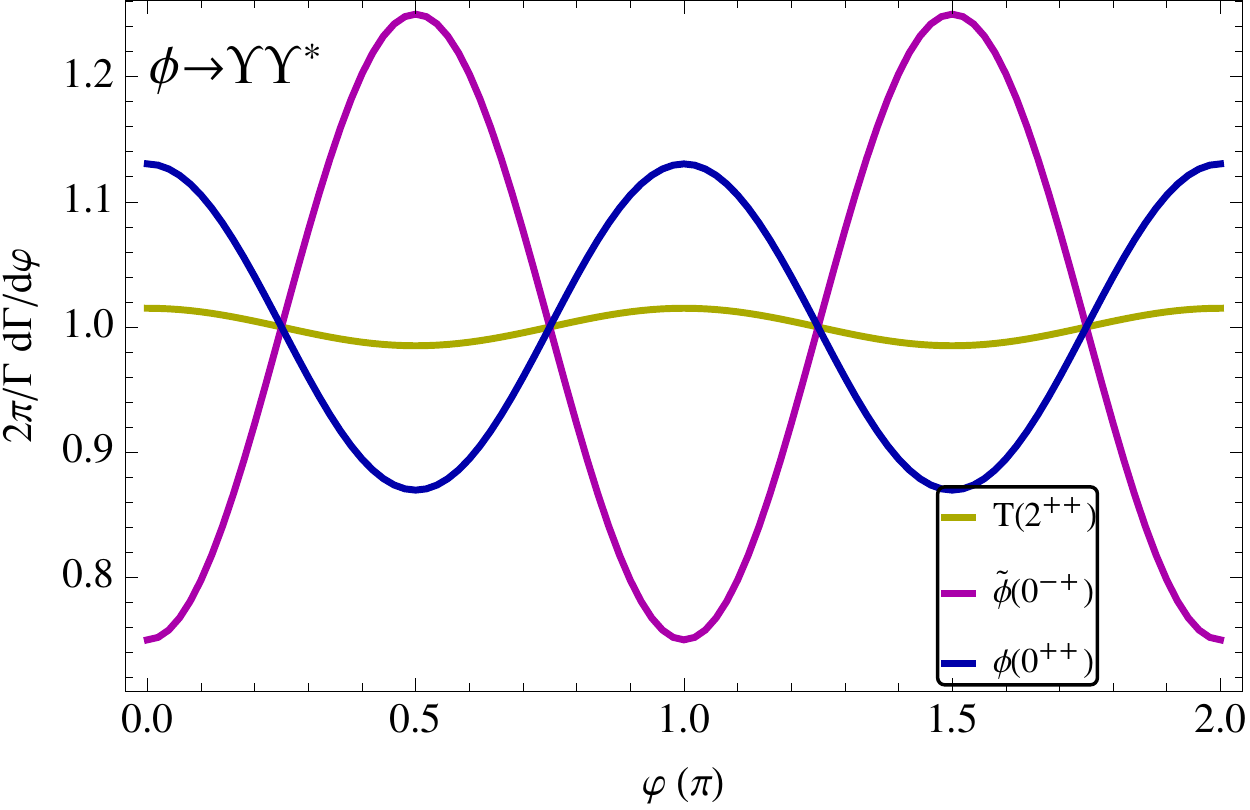}
  \caption{
  Upper panel: The angular distribution of the angles between the decay planes of the $\Upsilon\Upsilon^*$ system for several benchmark values of the off-shell $\Upsilon$ masses from a $0^{++}$ state. 
  Lower panel: The angular distribution of the angles between the decay planes of the $\Upsilon\Upsilon^*$ system for various assumptions of the spin and parities of the resonant particle.
  }  
  \label{fig:decayangularphi}
  \end{center}
\end{figure}

For the tetraquark state decaying into four leptons final states, there are several important physics observables for us to understand the properties of this state. We discuss them in order. Assuming that the only observable decays are from $\phi\to \Upsilon (1S) \Upsilon(1S)^*$, we can derive many useful differential distributions that are informative in identifying the physics origin of this state, similar to the case of the SM Higgs decaying to four leptons through intermediate $Z$-bosons~\cite{Choi:2002jk}. 
We use the differential formalism detailed in Eq.(27) and include additional off-shell suppression and velocity suppression factors for the invariant mass distribution detailed in Eq.(23) of Ref~\cite{Choi:2002jk}. We note here that due to vector meson dominance, the axial vector terms proportional to $\eta_1$ and $\eta_2$ in these formulae are all zero. Furthermore, the coefficients of our interaction terms under consideration in Table.~\ref{tab:EFTs} can be identified as coefficients of $a_1$, $c_1$ and $a_1$ for the $0^{++}$, $2^{++}$ and $0^{-+}$ states, respectively, in Table. 1 of Ref.~\cite{Choi:2002jk} when matching the helicity components for the calculation.
To study the differential distributions of the tetraquark ground state, we choose the mass to be slightly below the $\eta_b\eta_b$ threshold, 18.5~GeV throughout the text unless noted otherwise.

In the upper panel of Fig.~\ref{fig:decayangular} we show the (polar) angular distributions of the di-lepton pairs in their corresponding rest frames. For a $0^{++}$ state, helicity component $T_{00}$, proportional to $\sin^2\theta_1 \sin^2\theta_2$, is independent from additional suppression in addition to the common off-shell propagator suppression. In contrast, the decay into the transverse vector states $T_{11}$, proportional to $(1+\cos^2\theta_1)(1+\cos^2\theta_2)$, are suppressed by additional factors. These behaviors result in the dependence of the angular distribution on the mass of the off-shell $\Upsilon$. 
In the lower panel of this figure, we show the angular distribution for various assumptions on spin and parity of the underlying resonant particle. For the $0^{-+}$ state, the interaction is dominated by the transversely polarized vector mesons and thus have a behavior favoring the forward and backward direction for the polar-angular distribution of the leptons. In contrast, for both the $0^{++}$ and $2^{++}$ case the transversely polarized intermediate vector mesons are suppressed and thus exhibit comparatively less angular dependence. 

In the upper panel of Fig.~\ref{fig:decayangularphi}, we show the angular distribution of the angles between the decay planes of the $\Upsilon\Upsilon^*$ system for several benchmark values of the off-shell $\Upsilon$ masses from a $0^{++}$ state. In the extreme case of the off-shell pair mass approaching zero, the distribution is completely flat, as the helicity $T_{11}$ component vanishes. In other cases, we can see distributions generated proportional to $\cos2\varphi$ superimposed on the flat distribution from helicity $T_{00}$ component. In the lower panel of this figure we show the angular distribution of the decay planes for various assumption on the spin and parity of the underlying resonant particle. The CP-odd state $0^{-+}$ features an opposite behavior in contrast to the case of the CP-even state of $0^{++}$ and $2^{++}$, as expected since this plane angle is a CP-sensitive observable.

\begin{figure}[htbp]
\begin{center}
\includegraphics[scale=0.7,clip]{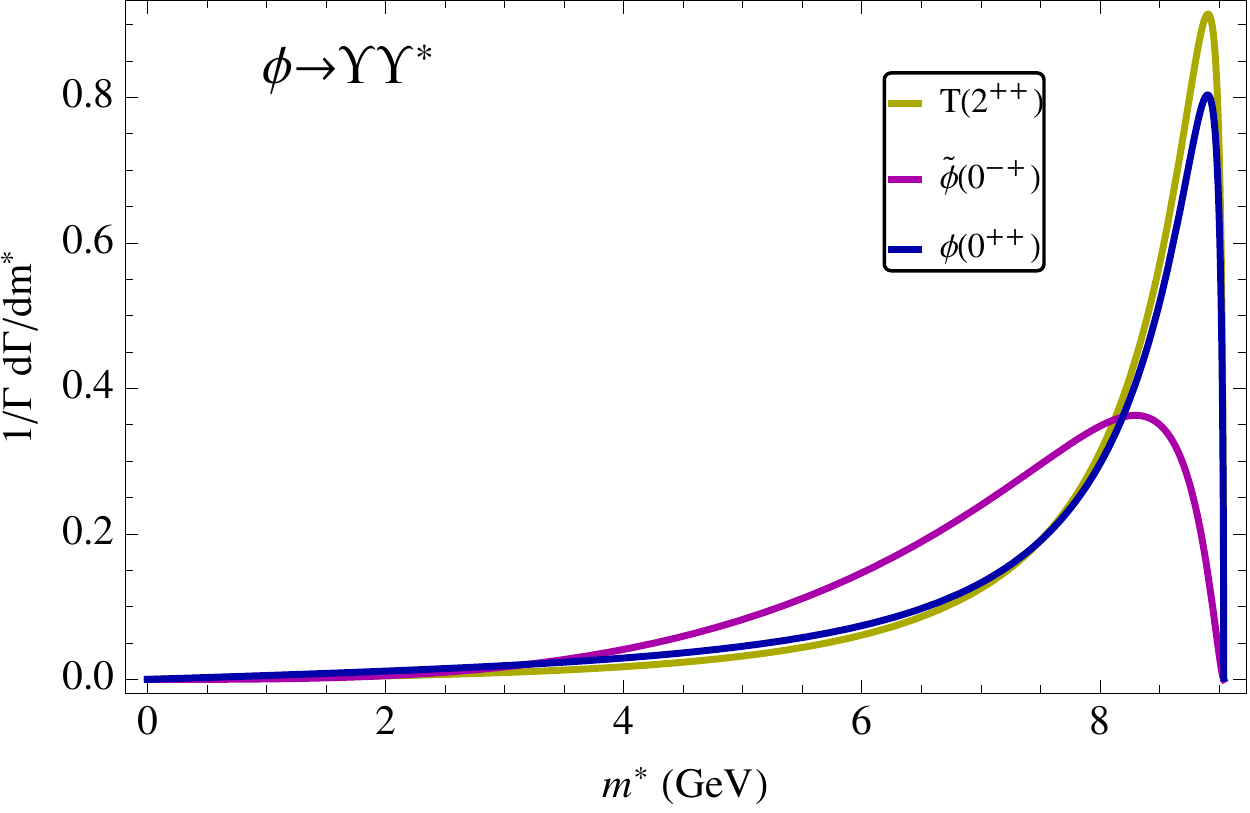}
\caption{
The normalized off-shell $\Upsilon(1S)$ invariant mass distribution for the process of $\phi\to \Upsilon(1S) \Upsilon(1S)^*$ for a $0^{++}$, $0^{-+}$ and $2^{++}$ hypothesis of the $\phi$ shown in blue, magenta and gold lines, respectively.
}  
\label{fig:invariantmass}
\end{center}
\end{figure}

In Fig.~\ref{fig:invariantmass} we show the normalized off-shell $\Upsilon(1S)$ invariant mass distribution for the process of $\phi\to \Upsilon(1S) \Upsilon(1S)^*$. 
As anticipated from the off-shell suppression behavior, the off-shell dilepton system is inclined to have high invariant mass to minimize the off-shell suppression. 
Still, different underlying state results in different behaviours in details of the off-shell dilepton invariant mass distribution. We can see the $0^{++}$ and $2^{++}$ hypothesis provide the sharply peaked invariant mass distribution because they are  $s$-wave processes, while the $0^{-+}$ hypothesis is $p$-wave suppressed. 


\subsection{Production properties} 

The production of the tetraquark state at the LHC will be mainly from the $gg$ initial state and with subsequent splitting into heavy quark pairs. These heavy quark pairs then form the color singlet states $\Upsilon$, $\eta_b$ and color octet states $\Upsilon^8$, $\eta_b^8$. When having low relative momentum, these pairs of bottomonium states can form the tetraquark state. For simplicity, we focus on the contribution from the $\Upsilon\Upsilon$ state for the production, which provide a conservative estimate of the production rate for the tetraquark state. The total inclusive production rate can then be expressed as
\bea
\sigma(pp\to \phi)&=&\int_{\tau_{\rm min}}^{\tau_{\rm max}} d\tau \frac {d \mathcal {L}} {s~d \tau} \hat \sigma (gg\to \Upsilon\Upsilon)\frac{dPS1} {dPS2}  |\bra{\Upsilon\Upsilon}\ket{\phi}|^2\\
&=& \int_{\tau_{\rm min}}^{\tau_{\rm max}} d\tau \frac {d \mathcal {L}} {s~d \tau} \hat \sigma (gg\to \Upsilon\Upsilon)\frac {8\pi \Lambda^2} {\tau s}
\label{eq:production} 
\eea
with
\beq
\frac {d\mathcal{L}} {d\tau} =f_g\otimes f_g (\tau)= \int_{\tau}^{1} \frac {dx} {x}  f_g(x;Q^2) f_g(\frac {\tau} x;Q^2),
\eeq
where $dPS1$ and $dPS2$ denote the one- and two-body phase space, respectively. The integration range $\tau_{min}\sim\tau_{max}$ is roughly determined by the sum of the masses of the constituent quarks to the $\eta_b\eta_b$ threshold, assuming partial sum rule to be valid. In this calculation, we adjust the pole mass of the $\Upsilon$ accordingly to the  bottom masses. 
A more rigorous treatment could be developed using the techniques of QCD sum rules applied to the
four current correlator~\cite{Wang:2017jtz} in an analogous way that finite energy sum rules have been employed to study the threshold region for heavy quark pair production in $e^+e^-$ collisions.

The partonic cross section for di-Upsilon production using s-wave production approximation can be found in Ref.~\cite{Li:2009ug} for color singlet pair productions and Ref.~\cite{Ko:2010xy} for color octet pair productions. We adopt these formulas for the partonic cross sections and convolute with {\tt NNPDF}~\cite{Ball:2012cx}. We reproduced their results and further verified our implementations for these double vector quarkonium production with current LHC measurements~\cite{Khachatryan:2014iia,CMS-PAS-BPH-14-008}. We obtain 39 fb for total inclusive double $\Upsilon$ production and 11 fb with rapidity cut of 2 on the final state $\Upsilon$s at 8 TeV LHC. Current experimental results are $68.8\pm12.7\pm 7.4\pm2.8$ pb~\cite{CMS-PAS-BPH-14-008}. For the $\Upsilon(1S)$ production, there will be contributions from double-parton scattering and from decays of higher excitation states, which will not contribute to the tetraquark ground state production. Hence, we use the matrix element squared obtained from Ref.~\cite{Li:2009ug} for our estimation of the cross section.
We further require the center of mass energy $\hat s \equiv \tau S$ in a window between 17.7~GeV to 18.8~GeV, where the lower bound is four times the bottom mass ($\overline {\rm MS}$) and the upper bound is the $\eta_b(1S)\eta_b(1S)$ threshold.

\begin{figure}[!htb]
  \begin{center}
  \includegraphics[scale=0.6,clip]{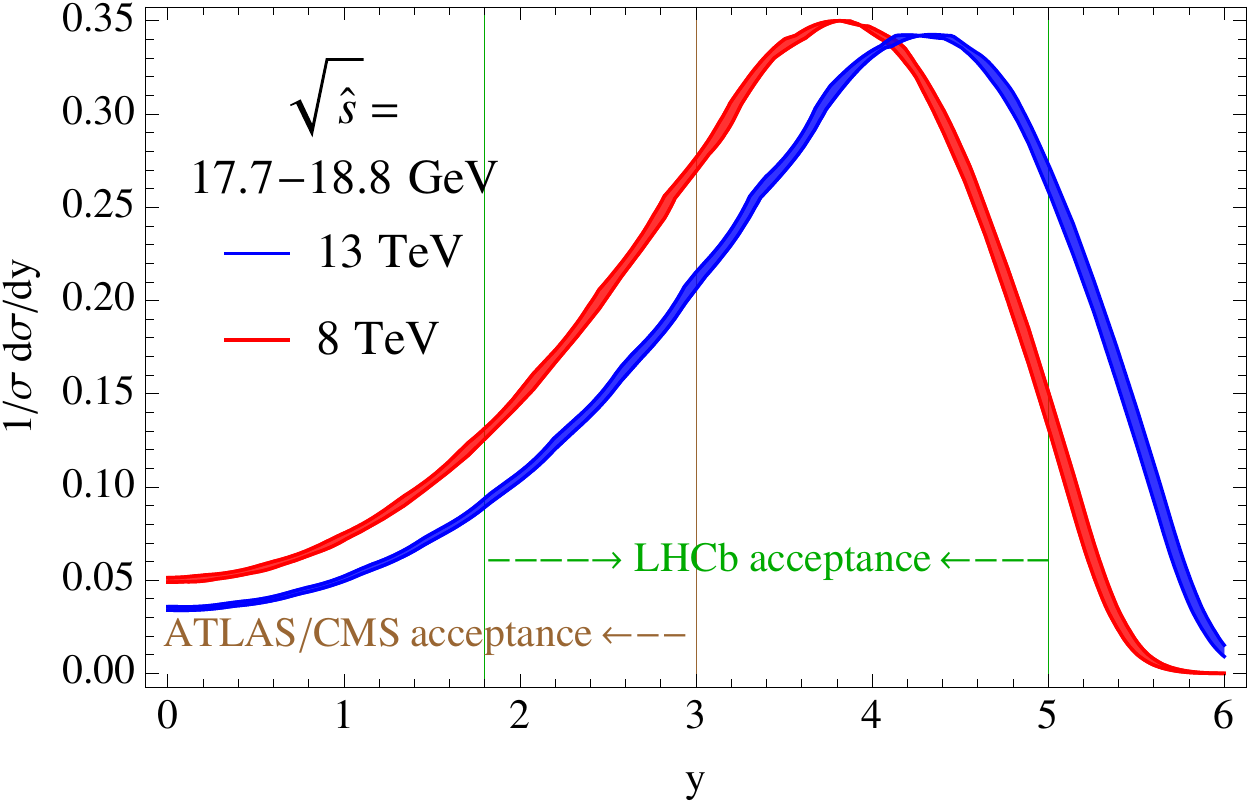}
  \caption{
  The rapidity distribution for this state at the LHC from glu-glu-fusion process for 8 TeV (red line) and 13 TeV (blue line) center of mass energy. 
  }  
  \label{fig:production}
  \end{center}
  \end{figure}

The decays branching fractions to four leptons near the threshold can be approximated by the square of the branching fraction of $\Upsilon(1S)$ to dileptons, BR($\Upsilon(1S)\rightarrow \ell^+\ell^-$)=4.9\%~\cite{Patrignani:2016xqp}. With above formalism for the production in Eq.~\ref{eq:production}, following our parameterization of the wavefunction of the tetraquark state in Eq.~\ref{eq:EFTs}, we can express the LHC production rate for such a ground state as 
\bea
\rm{8~TeV}:&&~\sigma(pp\rightarrow \phi \rightarrow 4\ell)\sim 3 \left(\frac {\Lambda} {0.2\gev}\right)^2~{\rm fb} \nonumber \\
\rm{13~TeV}:&&~\sigma(pp\rightarrow \phi \rightarrow 4\ell)\sim 5 \left(\frac {\Lambda} {0.2\gev}\right)^2~{\rm fb}. \nonumber
\label{eq:productionrate} 
\eea
The production rate for the whole process $\sigma(pp\to \phi\to \mu^+\mu^-\mu^+\mu^-)$ is then $\mathcal{O}(1)$ fb and also for $4e$ and doubled for $2\mu2e$ final states at 13 TeV LHC.  
However, the ground $0^{++}$ state is also anticipated to have sizable wavefunction overlaps with the $\eta_b\eta_b$ state, the production of which will further increase the cross section for the tetra-quark states.\footnote{Having this wavefunction overlap with $\eta_b\eta_b$ will change the decay partial width to four leptons as well.} Hence, the equation provided above can be viewed as an estimation for the typical production rate. 

In addition to the production rate information, the rapidity distribution of the signal events also can be useful as some consistency check. In Fig.~\ref{fig:production} we show the signal rapidity distribution for both LHC 8 TeV and 13 TeV, in red and blue lines, respectively. The tetraquark state turns to be produced with a peak rapidity of 3.8 (4.4) for LHC 8 (13) TeV, which would further impact the kinematic distribution of decay products. A general feature of the decay products should be having at least one dilepton pairs in the forward region. The bands in the figure indicates the rapidity distribution in variation according to the partonic center of mass energy that varies between 17.7~GeV and 18.8~GeV. We further draw band of typical acceptance in rapidity for ATLAS/CMS and LHCb. The LHCb forward coverage features a generically larger acceptance for the low-lying state produced through gluon-gluon fusion. However, given the relatively low cross section for this process and the lower luminosity accumulated by the LHCb, it would still be hard for such a state to be found in current data. However, the tetra-quark state is anticipated to decay into many other final states, at least following the behavior of pair-produced $\Upsilon(1S)$ states. LHCb may provide unique probe or discovery for such a state with their lower threshold designed to capture forward bottom hadrons. Despite that the acceptance should be applied to final state leptons instead of the resonant tetraquark state, the rapidity distribution does provide a testable property if this state is observed. Furthermore, since the rapidity behavior if driven by the gluon PDF, the distribution show in the right panel of Fig.~\ref{fig:production} would still hold for a generic resonant particle produced though the 
gluon-gluon-fusion process in the vicinity of the mass window under consideration. 


\section{Summary and outlook} 

In this paper we have investigated the observational details of detecting a bound tetra-bottom state with a mass below the di-$\eta_b(1S)$ threshold.  The ground state, $\phi$, would be very narrow and
likely would  have $J^{PC}=0^{++}$.  The most promising discovery mode at the LHC would be through the decay of $\phi$  into four charged leptons approximately  described by an effective interaction of the form $\Lambda \phi \Upsilon^{\mu}\Upsilon_{\mu}$.  Although decays involving  $\Upsilon (2S)$ and $\Upsilon(3S)$ might also be observed, the ground state of $\Upsilon(1S) \Upsilon(1S)^*$ should dominate. With this simple model, many properties of the possible low-lying tetraquark state can be tested.

We compute the expected angular distributions for the $\ell^+\ell^-$ arising from decays of the on-shell and off-shell $\Upsilon$ states as a function of the off-shell dilepton system invariant mass in Fig.~\ref{fig:decayangular}. Furthermore, an off-shell dilepton mass dependent angular correlations between the decay planes of the $\Upsilon\Upsilon^*$ system can be found in Fig.~\ref{fig:decayangularphi}. The off-shell dilepton invariant mass distribution should be peaked toward high invariant mass, as preferred by the off-shell $\Upsilon$ propagator. 
We show the angular distribution of the dilepton system, angular distributions between the $\Upsilon\Upsilon^*$ decay planes, and the invariant mass distributions with different underlying assumptions about the spin and CP property of the tetraquark state in (the lower panels of) Fig.~\ref{fig:decayangular}, Fig.~\ref{fig:decayangularphi} and Fig.~\ref{fig:invariantmass}, respectively. 

Furthermore, we estimated the possible cross section of the low-lying tetraquark state using partial sum rules, and found its cross section dependence on the model parameters, with a typical cross section of $O(\rm fb)$ for $\Lambda=0.2~\gev$. The rapidity distributions of such a tetra-bottom state should be dominated by the gluon-gluon-fusion process, which features a very forward behaviour due to the gluon PDF behavior. This behaviour provides testable predictions and interesting implications on the complementarity between ATLAS/CMS and LHCb experiment at the LHC. 

Finally we note,  that the decay angular distributions are generic for a massive state that couples to $\Upsilon$ plus a massive vector state in the mass range we consider, 
depending only on the $J^{PC}$ of the decaying state; and that the rapidity distribution of production only depends on the PDF behavior of the initial state gluons. 
Hence, most of our results would generically useful for testing low-lying states at the LHC.




{\it \bf Acknowledgments:} We thank Yang Bai, Kiel Howe, Ciaran Hughes 
for helpful discussion. 
This manuscript has been authored by Fermi Research Alliance, LLC under Contract No. DE-AC02-07CH11359 with the U.S. Department of Energy, Office of Science, Office of High Energy Physics. The United States Government retains and the publisher, by accepting the article for publication, acknowledges that the United States Government retains a non-exclusive, paid-up, irrevocable, world-wide license to publish or reproduce the published form of this manuscript, or allow others to do so, for United States Government purposes.

\bibliographystyle{JHEP}
\bibliography{fourQ}

\end{document}